# Data Analytics Approach to Predict High-Temperature Cyclic Oxidation Kinetics of NiCr-based Alloys[*]

Jian Peng, Rishi Pillai, Marie Romedenne, Bruce A. Pint, Govindarajan Muralidharan,

J. Allen Haynes, Dongwon Shin[†]

Materials Science and Technology Division, Oak Ridge National Laboratory, Oak Ridge, TN 37831

[†]Email address: shind@ornl.gov

**Abstract**: Although of practical importance, there is no established modeling framework to accurately predict high-temperature cyclic oxidation kinetics of multi-component alloys due to the inherent complexity. We present a data analytics approach to predict the oxidation rate constant of NiCr-based alloys as a function of composition and temperature with a highly consistent and well-curated experimental dataset. Two characteristic oxidation models, i.e., a simple parabolic law and a statistical cyclic-oxidation model, have been chosen to numerically represent the high-temperature oxidation kinetics of commercial and model NiCr-based alloys. We have successfully trained machine learning (ML) models using highly ranked key input features identified by correlation analysis to accurately predict experimental parabolic rate constants ($k_p$). This study demonstrates the potential of ML approaches to predict oxidation kinetics of alloys over a wide composition and temperature ranges. This approach can also serve as a basis for introducing more physically meaningful ML input features to predict the comprehensive cyclic oxidation behavior of multi-component high-temperature alloys with proper constraints based on the known underlying mechanisms.

[*]Notice: This manuscript has been authored by UT-Battelle, LLC, under contract DE-AC05-00OR22725 with the US Department of Energy (DOE). The US government retains and the publisher, by accepting the article for publication, acknowledges that the US government retains a nonexclusive, paid-up, irrevocable, worldwide license to publish or reproduce the published form of this manuscript, or allow others to do so, for US government purposes. DOE will provide public access to these results of federally sponsored research in accordance with the DOE Public Access Plan (http://energy.gov/downloads/doe-public-access-plan).

**INTRODUCTION**

Ni-based alloys used at high temperatures are required to have both good mechanical properties and oxidation resistance. These alloys obtain their oxidation resistance by forming adherent protective chromia and/or alumina scales[1,2]. Currently, the evaluation of alloys' high-temperature oxidation still heavily relies on experimental investigations, which is costly and time-consuming. Hence, developing the capability to predict the high-temperature oxidation kinetics of multi-component alloys is highly desirable and of great interest in many applications involving extreme environments. Although computational approaches using conventional analytical and physics-based simulations have made significant advancements in investigating high-temperature oxidation, they are usually only applicable for a given length/time scale[3-6] or specific oxidation-related phenomena[7-11].

On the other hand, improvements in the machine learning (ML) approach have propelled the use of data analytics to assist the discovery of new materials and the prediction of properties[12-19]. It possesses many advantages in handling complex multi-component alloys and offers the potential to extract insights from complex experiments or synthetic datasets generated from physics-based simulations. Thus, modern data analytics approaches can be considered an alternate and/or complementary tool to accelerate the design and development of novel materials with reduced cost and risk[12-14,20-22].

Recently, data analytics approaches have been successfully applied to predict the mechanical properties of multi-component high-temperature alloys[18,22-27]. While high-temperature oxidation also has scientific and practical importance, to the best of the authors' knowledge, very limited effort has been made to predict the oxidation kinetics of complex multi-component alloys by ML.

A preliminary attempt has been made by Bhattacharya et al.[28] to build ML models to predict the high-temperature oxidation kinetics, described by the parabolic rate constant ($k_p$), of Ti alloys between 550 and 750°C. From a statistical perspective, a good agreement between the predicted and experimental $k_p$ was achieved. However, their dataset was collected from multiple literature sources; thus, the data may exhibit mixed oxidation mechanisms due to the difference in experimental protocols (i.e., atmospheres and isothermal or cyclic exposures)[26]. Moreover, their work focused only on building predictive models without analyzing the relationship between input features and $k_p$.

[Figure 1 about here]

Herein, we present a new data analytics framework consisting of correlation analysis of a consistently measured and well-curated oxidation dataset followed by ML to predict the cyclic oxidation kinetics of NiCr-based alloys, as illustrated by the workflow shown in Figure 1. The oxidation kinetics of studied alloys was numerically represented by $k_p$ from two representative oxidation models. Correlation analysis was performed using two algorithms (Pearson's correlation coefficient (PCC)[29] and maximal information coefficient (MIC)[30] to quantitatively evaluate the strength of correlation between input features (i.e., alloying compositions/oxidation temperature) and $k_p$, and hence rank the input features. The performance of five widely used ML models, i.e., linear regression (LR)[31], Bayesian ridge (BR)[32,33], $k$-nearest neighbor (NN)[34], random forest (RF)[35], and support vector machines (SVM) regression[36] in accurately predicting the $k_p$ as a function of the number of top-ranking input features is evaluated and discussed in this paper.

**RESULTS AND DISCUSSION**

**Numerical Representation of Cyclic Oxidation Data**

Extensive experimental investigations of the cyclic oxidation kinetics of 25 model NiCr alloys and 4 commercial Ni-based alloys (i.e., Nimonic 80A, Nimonic 90, René 41, Haynes 282) have been performed in both dry air and wet air (air+10% water vapor) atmospheres at Oak Ridge National Laboratory (ORNL). Figure 2 shows examples of the cyclic oxidation behavior of select NiCr alloys[37,38] in the dataset. The alloy composition ranges, experimental conditions, and details of the dataset are summarized in Table 1. This consistently measured oxidation and well-curated dataset, collected over years of experiments at ORNL, enabled exploring the potential of ML to predict the oxidation kinetics of multi-component high-temperature alloys. However, raw cyclic oxidation data, which is given as a 2×$N$ matrix (cycle vs. mass change) per alloy, cannot be directly used in ML because singular numeric values are needed as the target for the regression-type ML training. Hence, it is necessary to find a proper numerical representation for the oxidation kinetics of each alloy so that the correlation between the input features and the target, as well as the capability of ML in modeling the high-temperature oxidation kinetics, can be evaluated.

[Figure 2 about here]

[Table 1 about here]

High-temperature oxidation of multi-component alloys is a complex dynamic process, which requires concurrent consideration of both thermodynamics and diffusion kinetics[2]. As shown in Figure 2, the addition of certain minor alloying elements can significantly alter the mass change behavior of some alloys during oxidation. Thus, identifying a proper numerical representation for such complex phenomena is the key to building reliable ML models to predict the high-temperature oxidation kinetics of NiCr-based alloys. Mathematically, any mass change curve can be fitted with an arbitrary polynomial. However, parameters obtained from such numerical regression without physical meaning will not allow gaining insights into the oxidation mechanisms and compare alloy oxidation performance. Thus, we have focused on identifying widely accepted oxidation models with a robust formalism to determine physically meaningful parameters to represent the oxidation kinetics of the studied alloys.

As illustrated in Figure 3, two oxidation models, i.e., a simple parabolic law ($\Delta m=\sqrt{k_p t}$, where $\Delta m$ is the mass gain and $t$ is the oxidation time; s-$k_p$ hereafter)[39,40], and a statistical cyclic-oxidation model (p-$k_p$ hereafter)[41], were adopted in this work. Both models have been widely used to interpret the oxidation kinetics of numerous alloys[42-48]. In this study, the s-$k_p$ model evaluated the mass change data up to 100 h to exclusively represent growth rates with the minimized influence of mass loss. The p-$k_p$ model[41], consisting of two parameters, i.e., $p$ (discrete oxide spallation probability) and $k_p$, evaluated the complete curve, thus, enabling us to assess the capability of the ML approach to predict oxidation kinetics involving both oxide growth and loss processes. It is anticipated that the derived $k_p$ will be a good numerical representation of our raw experimental cyclic oxidation data and can serve as a proper target property for analyzing the oxidation kinetics by the data analytics approach. Since the oxidation mechanisms of NiCr-based alloys in dry air and wet air are different[49,50], the complete experimental data after being fitted by each oxidation model were divided into two datasets based on the oxidation atmosphere. Finally, four datasets, s-$k_p$ and p-$k_p$ in dry and wet air, were generated from the existing experimental data for further analysis in this study (see Table 1).

[Figure 3 about here]

**Correlation analyses**

We have used MIC and PCC methods to quantitatively analyze the correlation between all input features (alloy compositions and oxidation temperature from Table 1) and $k_p$, respectively. While PCC can evaluate the strength of the positive and negative, it only identifies linear relationships between two variables. The MIC method can determine the strength of both linear and non-linear relationships, but only the magnitude without a sign. Both approaches are expected to provide insights into the correlation between input features and $k_p$ from differing statistical aspects, which may inspire alloy design experts to generate new alloy hypotheses[23,26]. Moreover, correlation analysis can facilitate the training of high-fidelity ML models using highly ranked features. For the PCC analysis, a positive correlation between a feature and $k_p$ here implies that the increase of the feature will increase $k_p$, i.e., a higher oxidation rate, and vice versa. Figure 4 shows the PCC analysis results for the $p$-$k_p$ of dry and wet air datasets as examples. The oxidation temperature (T) was identified as the feature having the most substantial impact on $k_p$ by both PCC and MIC analyses. It is encouraging that correlation analysis replicated the community's existing knowledge: oxidation temperature is typically the factor that most strongly affects oxidation rate. PCC also correctly identified that temperature has the most positive correlation with $k_p$, consistent with the fact that the oxidation rate increases with increasing temperature.

It can also be observed in Figure 4a that chromium (Cr) as the major alloying element in the studied alloys was identified as a high-ranking feature, which is consistent with the fact that the Cr addition is beneficial to the formation of an external chromia solid-state diffusion barrier that slows the oxidation reaction as it increases in thickness[51,52]. Most of the alloys in this dataset were designed to form protective chromia scales. Aluminum (Al) is expected to further decelerate the growth rate of the chromia scale for NiCr-based alloys[53]; thus, the negative correlation between Al and $k_p$ in wet air (Figure 4b) is reasonable. Ni, Mn, and Fe are known to form fast-growing oxides[54-56]. The positive correlations between Ni and $k_p$ in dry air (Figure 4a), between Mn, Fe, Ni, and $k_p$ in wet air (Figure 4b), are in accordance with this understanding. Similar correlations were also observed for Cr in the dry air ($s$-$k_p$) dataset and Cr, Al, Ni, Fe in the wet air ($s$-$k_p$) dataset, as shown in Supplementary Figure S1. Other than identifying linear correlation as PCC does, MIC can also identify non-linear correlations. As presented in Supplementary Figure S2, features like T, Fe, Cr, Mo, Mn, and Ni in dry air and T, Al, Fe, and Cr in wet air that have a significant impact on the oxidation kinetics of NiCr alloys were generally identified as high-ranking features by the MIC analysis.

[Figure 4 about here]

Correlations that seem to contradict known oxidation mechanisms were also observed. For example, Fe in dry air datasets (Figures 4a and S1a) exhibits a negative correlation with $k_p$, meaning that Fe addition will decelerate the oxidation rate. Cr in the wet air ($p$-$k_p$) dataset (Figure 4b) shows almost no correlation with $k_p$. However, it should be emphasized that the rankings from correlation analysis here do not necessarily all make intuitive sense and mean causation. An unexpected correlation was also identified in a previous data analytics study on alumina-forming austenitic (AFA) alloys[23]. The negative dependence of Cu on creep properties was in contradiction because Cu addition was found to promote creep resistance[57,58]. This correlation discrepancy resulted from the fact that Cu was only added to AFA alloys with lower creep resistance, and correlation analysis might have "correctly" captured such inhomogeneity of the dataset[23].

High-temperature oxidation kinetics of multi-component alloys is controlled by chemical compositions and temperature and by factors including, but not limited to, microstructure and its evolution during oxidation, as well as oxide scale spallation/evaporation[2]. Correlation analysis results also heavily depend on the nature of the dataset. Although our dataset was collected over a number of years from differing studies at ORNL and could be considered a "big" dataset from a material science/alloy oxidation perspective, it is still small from a data science perspective. Thus, the dataset does not consistently cover the ideal composition ranges and test conditions to fully facilitate data analytics for the classes of alloys included. Certainly, this limitation could most likely be said of any existing materials dataset which includes complex alloy compositions, structures and behavior. However, the correctly identified correlation in Figures 4 and S1 still clearly demonstrated the value of correlation analysis in applying data analytics to materials science. The intention of performing correlation analysis here is not only to gain insights into the impact of input features on $k_p$ but also to have a numerical basis for the selection of input features to train ML models for understanding their influence on the performance of ML models. The remaining correlation analysis results are presented in Supplementary Figures S1 and S2.

**Machine learning**

Based on the ranking of features from correlation analyses, five widely used ML models, i.e., LR, BR, NN, RF, and SVM, have been trained to evaluate their performances. An introduction of these models is available elsewhere[59,60]. For models trained with the dry and wet air $p$-$k_p$ datasets, their average performance, represented by the Nash and Sutcliffe coefficient of efficiency (NSE)[61], and corresponding standard deviation as a function of the numbers of top-ranking features from the PCC analysis are shown in Figure 5. The remaining results are presented in Supplementary

Figures S3 and S4. Since NSE alone could not fully qualify the performance of these models, another metric, i.e., coefficient of determination ($R^2$ COD), and a band showing the goodness-of-fit were adopted to facilitate the assessment of ML models, as presented in Figure 6.

NSE here was mainly used to compare the performance of these models. As shown in Figures 5 and S3, the performance of ML models trained with dry air datasets shows a similar trend to those with wet air datasets. In most cases, increasing the top-ranking features from 2 to ~4 or ~6 increased the NSE of these models markedly. Intriguingly, considering more features did not considerably improve the performance of the obtained models. The NSE of the models trained with the wet air datasets is relatively lower than their dry air datasets counterparts. This trend can be attributed to the increased variation of measured oxidation kinetics of alloys studied in wet air exposure. For the dry air dataset, the maximum NSE of all models is between 0.6 and 0.75. Among the considered models, SVM has the highest NSE (>0.7) in all cases (Figures 5 and Supplementary Figures S3 and S4). It reaches the maximum NSE when using the top 3 to 5 features. Afterward, it gradually decreases with an increasing number of features, revealing that in this study including features with a ranking lower than the $5^{th}$ is not beneficial to the performance of ML models. For the wet air dataset, the maximum NSE of all models is only between 0.45 and 0.60. In general, the top 4 to 6 features in this study were required to achieve an NSE of 0.5 or higher. SVM still exhibits the highest accuracy among considered models. As discussed in the previous section, most of these features are experimentally identified features that significantly impact the oxidation of NiCr-based alloys.

[Figure 5 about here]

In all cases, the performance of these ML models evaluated with NSE is generally in the order of SVM>BR>RF≈NN>LR (in dry air) and SVM>BR≈LR>NN>RF (in wet air). Results show that these models are sensitive to the number of features considered in training but to different extents. SVM performs similarly to the linear-based models BR and LR, particularly for those trained with the wet air datasets. A linear kernel function (assuming the input features and $k_p$ is in a linear relationship) was adopted for the present SVM models after hyperparameter tuning. Considering more than the top 6 to 8 features significantly degraded the performance of LR for the dry air dataset, whereas this was not the case for the wet air dataset. This could be caused by the smaller data volume of the dry air dataset than the wet air dataset. BR uses probability distributors to formulate linear regression, rather than point estimates used by LR and identifies the posterior distribution for the model parameters. Thus, it is believed that BR is more tolerant of overfitting[32,33]. This is why, in this case, the NSE of the BR model does not notably decrease with an increasing number of features. RF in this study generally requires the inclusion of the top 4 features to obtain the NSE of >0.65 (for dry air datasets) and >0.5 (for wet air datasets); after that, its performance was almost independent of the number of considered features.

The observed trend can be understood as follows: as an ensemble learning method, RF assigns different importance to each feature during model training; thus, less critical features would have less or even no contribution to its performance[62]. Therefore, the performance of RF is not sensitive to the number of features considered in training after it reaches the maximum accuracy. NN hinges on the assumption that similar things are close to each other and simply outputs the average value of data points in $k$-nearest neighbors[60]. The performance of NN usually degrades as the dimensionality of data increases significantly[63], because its distance measure, i.e., the Euclidean distance, becomes less representative in a higher-dimensional space. However, as shown in Figure 5, Supplementary Figures S3 and S4, when the number of features increases, accompanying the increase of the dimensionality of data, the accuracy of NN does not decrease significantly. This finding indicates that the dimensionality of data in this study does not induce a significant adverse effect on the performance of NN. Since SVM possesses the best performance among all considered ML models, its performance does not significantly degrade with increasing numbers of features.

[Figure 6 about here]

[Figure 7 about here]

SVM models trained with all 11 features in our datasets are believed to be the most promising ML models for predicting the $k_p$ of NiCr-based alloys since they exhibit relatively good performance and concurrently consider essential features from a high-temperature oxidation perspective. Comparisons between the experimental and predicted $k_p$ by the SVM models trained with all 11 features are presented in Figures 6 (for $p$-$k_p$) and Supplementary Figure S5 (for s-$k_p$). While the accuracy of trained ML models evaluated with NSE is around 0.7 (not necessarily an impressively high value), it is encouraging that most of the predicted values are within the acceptable deviation range. In addition, the other metric $R^2$ COD for both dry and wet air ($p$-$k_p$) is close to 0.9 or higher, which strongly indicates that our ML models can efficiently capture the trend of cyclic oxidation response of NiCr-based alloys as a function of alloy compositions and oxidation temperature.

In practice, the same alloy tested under the same condition may exhibit significant test-to-test variations, particularly in wet air testing. For example, the $k_p$ from the $p$-$k_p$ model of Nimonic 80A at 950 °C in wet air from six tests varied significantly (Figure 7). Thus, dashed lines, as the acceptable deviation, representing one order of magnitude difference between the predicted and experimental $k_p$ are superimposed in Figures 6 and S5. The predicted $k_p$ correctly captures the experimental data trend, and most of the data points lie between the dashed lines. This indicates good agreement between experimental and predicted $k_p$ was achieved from a high-temperature oxidation perspective for the multi-component commercial alloys (group I). This conclusion is further supported by the high $R^2$ COD of $\geq 0.9$ in both cases. Data in Figure 6b are slightly more scatted than those in Figure 6a, attributing to the increased spallation probability and thus more considerable test-to-test variation in wet air exposure.

[Figure 8 about here]

Although a satisfactory agreement between the experimental and predicted $k_p$ was achieved, more work is required in the future. Firstly, the performance of ML models depends not only on what features have been considered but also on the nature and repartition of the dataset used for training; thus, further detailed analysis of this aspect is required. Secondly, $k_p$ alone could not fully capture the complete oxidation kinetics of alloys. Other properties, such as spallation probability and oxidation lifetime, should also be concurrently considered as target properties to be modeled/predicted. Thirdly, besides the alloy composition and temperature considered in this study as input features, the high-temperature oxidation kinetics are also affected by factors like microstructure, oxide scale spallation, alloy surface depletion of elements that are selectively oxidizing, and oxide evaporation in wet air atmosphere. Features that can well represent these additional factors and ML frameworks that can incorporate these factors as much as possible will be highly conducive to establishing high-fidelity surrogate models.

Additionally, material datasets commonly suffer from uneven data distribution. While the present dataset can be regarded as one of the largest and the most comprehensive cyclic oxidation dataset for high-temperature NiCr-based alloys, many gaps are identified. A highly desirable alloy dataset to apply data analytics would have an even distribution of data points of all elements without major gaps in a high-dimensional space. The movement toward such an ideal alloy dataset can be obtained by exercising a design of experiments (DOE)[64] campaign to fill key gaps in the existing dataset(s). However, the dataset used in the current study is biased to some extent and has a number of gaps, as shown in Figure 8. This is attributed to the fact that the objective of past empirical alloy research has been mainly focused on identifying alloys with superior properties over existing ones. Consequently, the alloy composition variation strategy has been heavily geared toward improving properties, more often ending up in biased and narrow boundary conditions from a data analytics perspective. It is evident that alloy data with poor properties can serve as equally useful learning input for ML models; however, an effort to strategically collect such data to fill the gaps has been overlooked, not only by us but likely throughout the materials community. Hence, a strategy that can identify the most important gaps in an existing dataset for alloy data analytics and efficiently augment a small amount of new experimental data needs to be developed to leverage legacy data, particularly for complex materials.

In summary, we demonstrated applying a data analytics approach to predict the cyclic oxidation kinetics of NiCr-based alloys using a highly consistent and well-curated experimental dataset collected over many years and numerous different studies at ORNL. The oxidation data in dry and wet air as a function of alloy compositions and oxidation temperatures (800~950°C) were used to train ML models. Identifying a proper numerical representation was the critical procedure for successfully predicting the high-temperature oxidation kinetics of NiCr-based alloys. Oxidation kinetics of alloys was represented by the parabolic rate constant ($k_p$) from properly selected oxidation models, i.e., a simple parabolic growth law for data up to 100 h (s-$k_p$) and a statistical cyclic-oxidation model for the complete curve (p-$k_p$, up to 2000 h), respectively.

Correlation analysis and training of ML models (i.e., BR, LR, NN, RF, and SVM regression) were performed to quantitatively evaluate the interrelationship between input features and $k_p$. The influence of top-ranking features on the performance of considered ML models was also identified. The trained ML models were able to predict $k_p$ from both s-$k_p$ and p-$k_p$ models satisfactorily. The typical test-to-test variation in experimental high-temperature oxidation data was believed to be the primary source for the discrepancy between the ML-predicted and experimental $k_p$. The current results demonstrated the potential of the ML approach to predict the oxidation kinetics of multi-component NiCr alloys with wide composition and temperature ranges. We anticipate that this newly demonstrated data analytics method can be applied to predict the oxidation kinetics of other classes of alloys.

**METHODS**

Experimental procedure

All specimens with the dimensions of ~10×20×1.5 mm$^3$ were ground to a 600-grit finish and cleaned ultrasonically in acetone and methanol prior to cyclic oxidation experiments. A 1 h-cycle consisting of a 60 min exposure at temperature and 10 min cooling in automated cyclic rigs[65] was adopted in this work. The specimens were exposed at temperatures between 800 and 950 °C in flowing dry air and wet air (air+10% H$_2$O), respectively, with a gas flow rate of 500 cm$^3$·min$^{-1}$ for up to 2000 h.

Correlation analysis and machine learning

The correlation between the input features and $k_p$ was evaluated by PCC[66] and MIC[30], respectively. PCC considers the strength of the linear relationship between two variables, while MIC can identify the strength of both linear and non-linear relationships. The correlation coefficient of PCC is between -1 and 1, where 1 indicates a total positive linear correlation, -1 represents a complete negative/reciprocal linear correlation, and 0 indicates no correlation. The correlation coefficient of MIC ranges between 0 and 1. Coefficient values close to 1 or -1 indicate the strongest correlation between the variables and target property.

Five representative ML models were used: LR[31], BR[32,33], NN[34], RF[35], and SVM regression[36]. These models were trained with different numbers of top-ranking features based on the ranking from MIC or |PCC| (the absolute value of PCC) to explore the influence of these features on the performance of ML models. The hyperparameters for each model were tuned up to 10,000 iterations using the randomized parameter optimization approach to search for the optimal parameters. Each model was trained 10 times for a given set of features to determine the averaged accuracy and its standard deviation. All features have the same weight in ML training regardless of their rankings determined by correlation analysis. Both correlation analysis and training of ML models were performed with the Python-based open-source frontend, Advanced data SCiEnce toolkit for Non-Data Scientists (ASCENDS)[60,67], which is available via GitHub (https://github.com/ornlpmcp/ASCENDS) The performance/accuracy of the ML models was quantified by NSE[61] and $R^2$(COD) calculated without an intercept as a supplementary metric. The NSE is calculated as follows:

$$NSE = 1 - \frac{\sum_{i=1}^{N}(O_i - P_i)^2}{\sum_{i=1}^{N}(O_i - \bar{O})^2}$$

where $O_i$ and $\bar{O}$ are the observed values and their mean, respectively. $P_i$ is the model predicted value. N is the number of data in the dataset. The *k*-fold cross-validation approach[68] with *k* = 5 was used in ML training. This approach randomly divided the input data into *k* groups. One group (i.e., unseen data) was withheld during training, and the remainder *k*-1 groups were used to train the ML model. Then the unseen dataset was used to evaluate the accuracy of models.

ACKNOWLEDGEMENTS

This research was sponsored by the Department of Energy, Energy Efficiency and Renewable Energy, Vehicle Technologies Office, Propulsion Materials Program. This research used resources of the Compute and Data Environment for Science (CADES) at the Oak Ridge National Laboratory, which is supported by the Office of Science of the U.S. Department of Energy under Contract No. DE-AC05-00OR22725. The authors thank C. Layton for his support for using CADES resources and G. Garner for assisting the oxidation experiments.

DATA AVAILABILITY

The data that support the findings of this study are available from the corresponding authors upon reasonable request.

AUTHOR CONTRIBUTIONS

D.S. conceived the study. R.P., M.R. and B.P. curated and prepared the dataset. J.P. performed correlation analysis and machine learning training. Commercial alloys relevant to exhaust valve environment were selected by G.M, B. P. and J. A. H. All authors analyzed the data. J.P. and D.S. drafted the manuscript. All authors reviewed the manuscript.

**COMPETING INTERESTS**

The authors declare no competing interests.

Table 1 List of alloy composition ranges in wt.% and experimental conditions, and details of the datasets

| | |
|---|---|
| Compositions of model alloys | Ni (41.7-86), Cr (14-25), Al (0-3.9), Co (0-16.2), Fe (0-40.1), Mn (0-1.5), Mo (0-9.8), Si (0-0.5), Ti (0-3.3), Y (0-0.1) |
| Commercial alloys | Haynes 282 (H282), Nimonic 80A (N80), Nimonic 90 (N90), Rene 41 (R41) |
| Temperature, °C | 800, 850, 900, 950 |
| Atmospheres | Dry air and wet air (air+10% water vapor) |
| Cycles of oxidation | Up to 2,000 cycles (1h/cycle) |
| Number of data points in each dataset | dry air (s-$k_p$): 58, dry air (p-$k_p$): 58<br>wet air (s-$k_p$): 151, wet air (p-$k_p$): 151 |
| Input features | [elements]: Elemental composition in wt.%<br>T: Oxidation temperature |
| Target | Parabolic rate constant ($k_p$) |

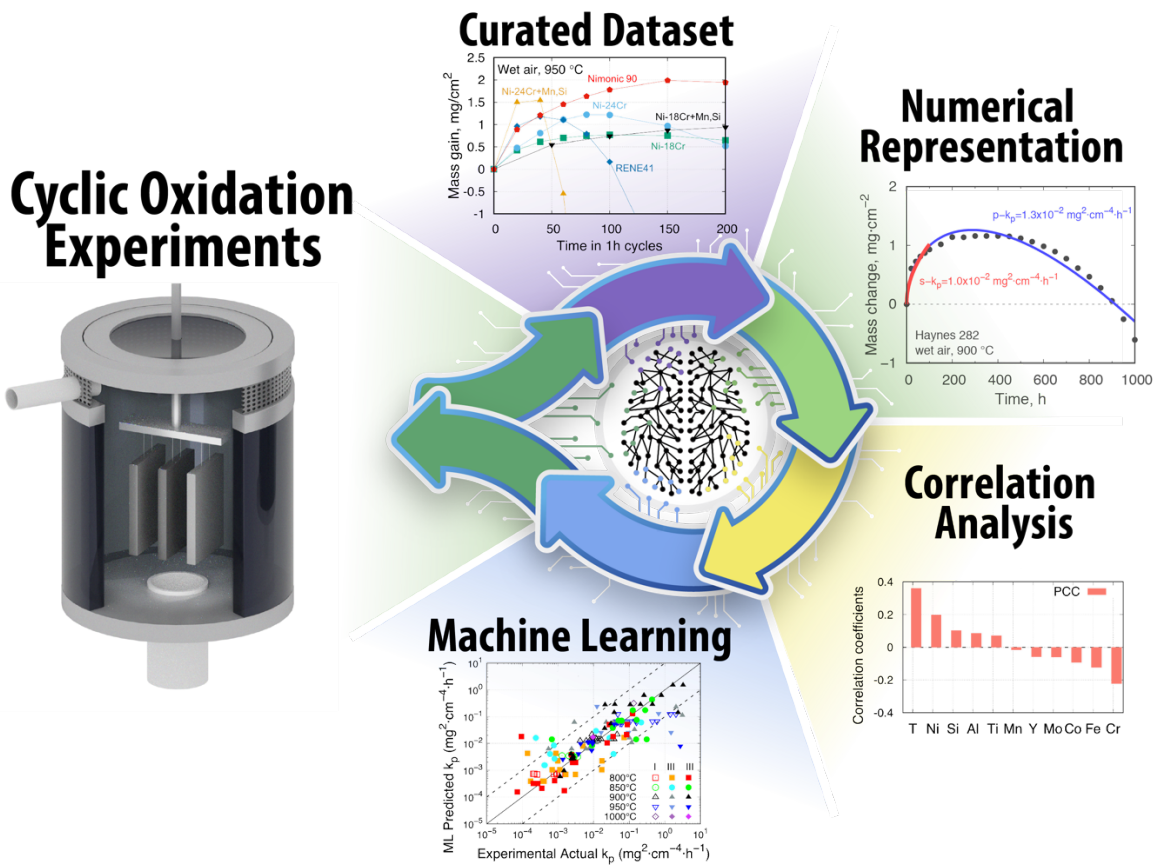

Figure 1 **Workflow of the present study.** Raw experimental data, mass gain curves from cyclic oxidation experiments, were collected and fitted with oxidation models to represent their oxidation kinetics by $k_p$ resulting in a dataset consisting alloy compositions, oxidation temperatures and $k_p$. Next, correlation analysis was performed to evaluate the strength of correlation between input features (i.e., alloy compositions and oxidation temperatures) with the target property ($k_p$). Then, based on the determined correlation coefficients, the input features were ranked and selectively used to train ML models. The performances of various ML models were evaluated.

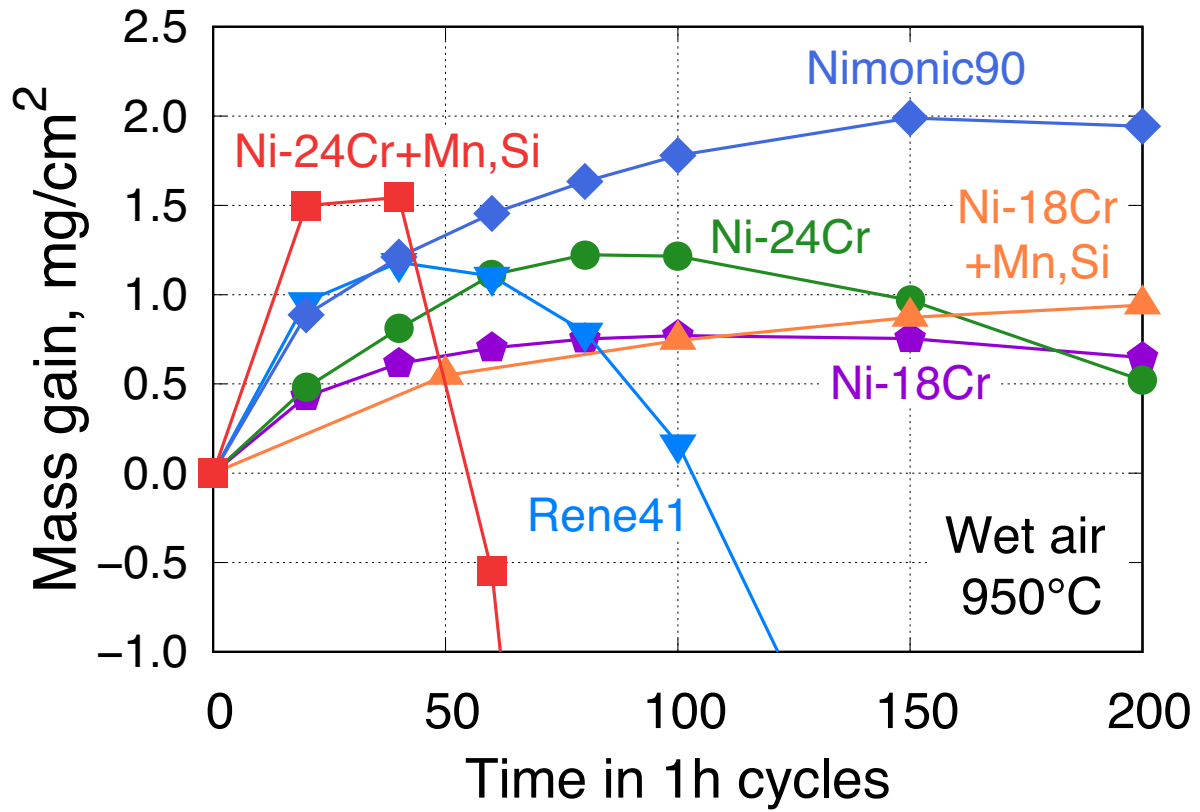

Figure 2 **Snapshot of ORNL cyclic oxidation dataset.** Mass gain of select alloys as a function of oxidation time in 1 hour cycles. The cyclic oxidation of alloys can be significantly altered by minor alloying elements.

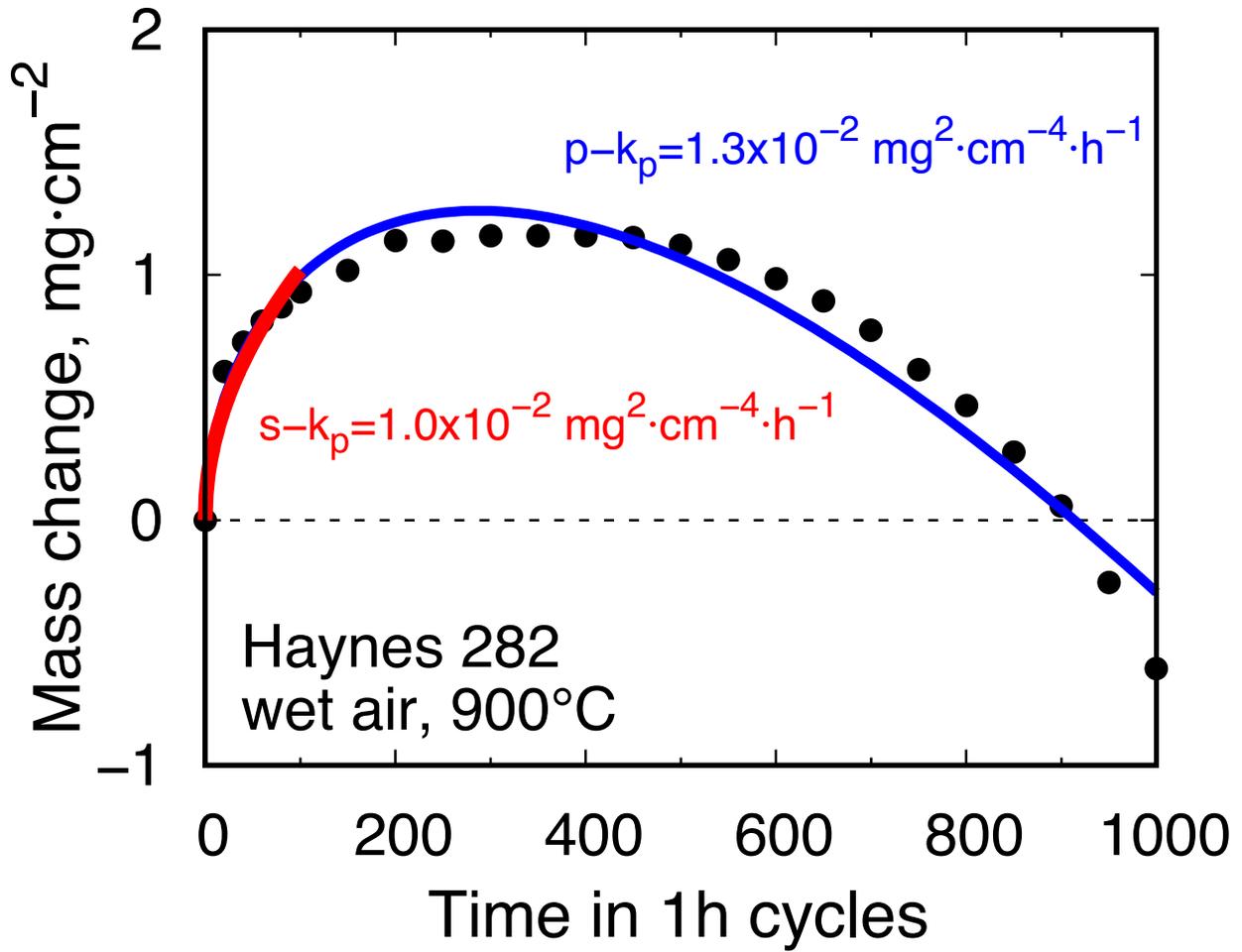

Figure 3 **Schematic diagram of fitted mass change curve.** The mass gain curve of Haynes 282 in wet air at 900 °C in 1 hour cycle was fitted with the s-$k_p$ and p-$k_p$ models, respectively.

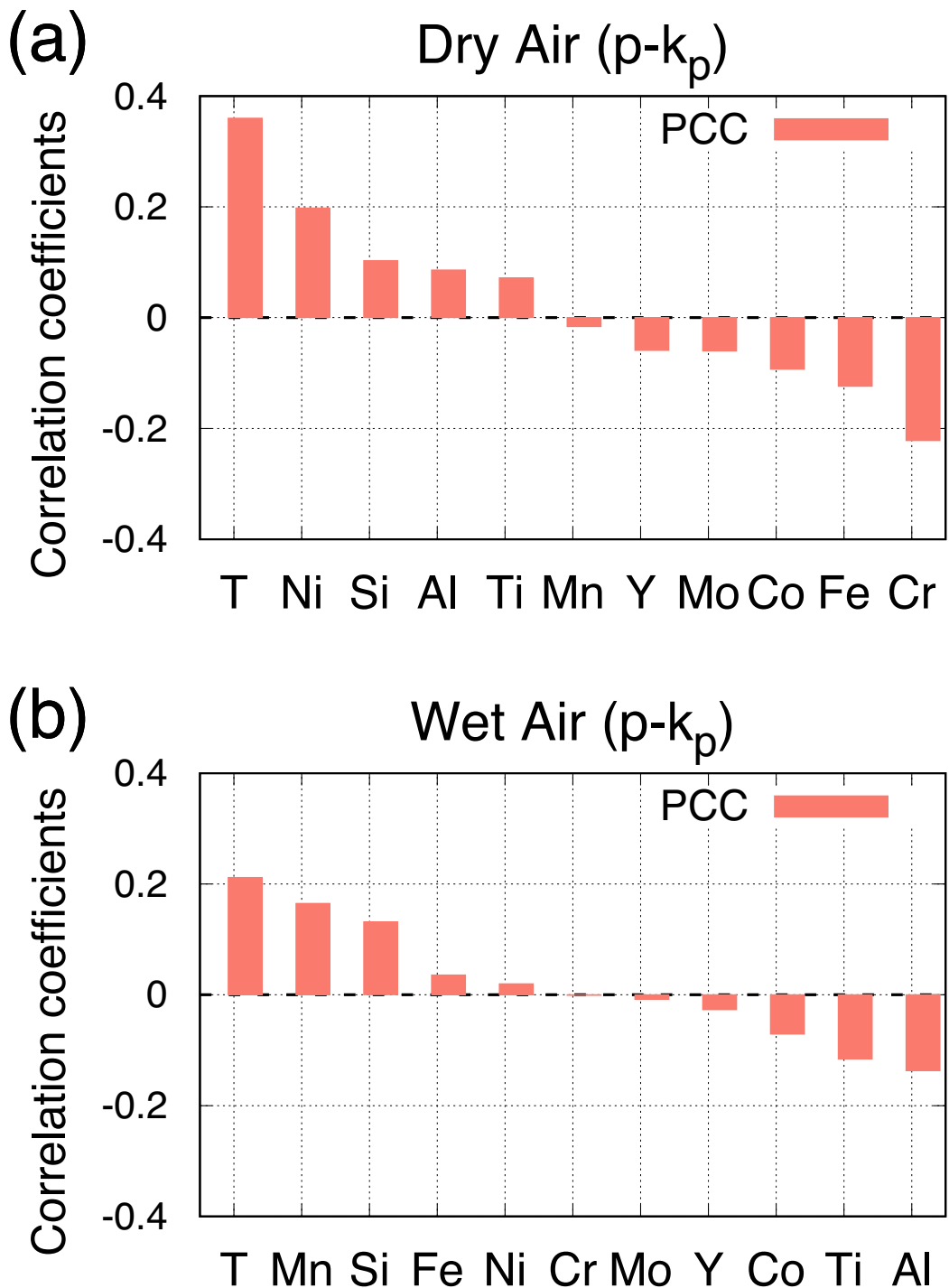

Figure 4 **Correlation analysis.** Quantified correlation scores of input features (elemental compositions and oxidation temperature) and $k_p$ determined for the s-$k_p$ and p-$k_p$ models, respectively, using Pearson correlatioin coefficient at difference atmpsphere. |PCC| denotes the absolute value of PCC coefficients. Details are reported in the Supplementary Table S1.

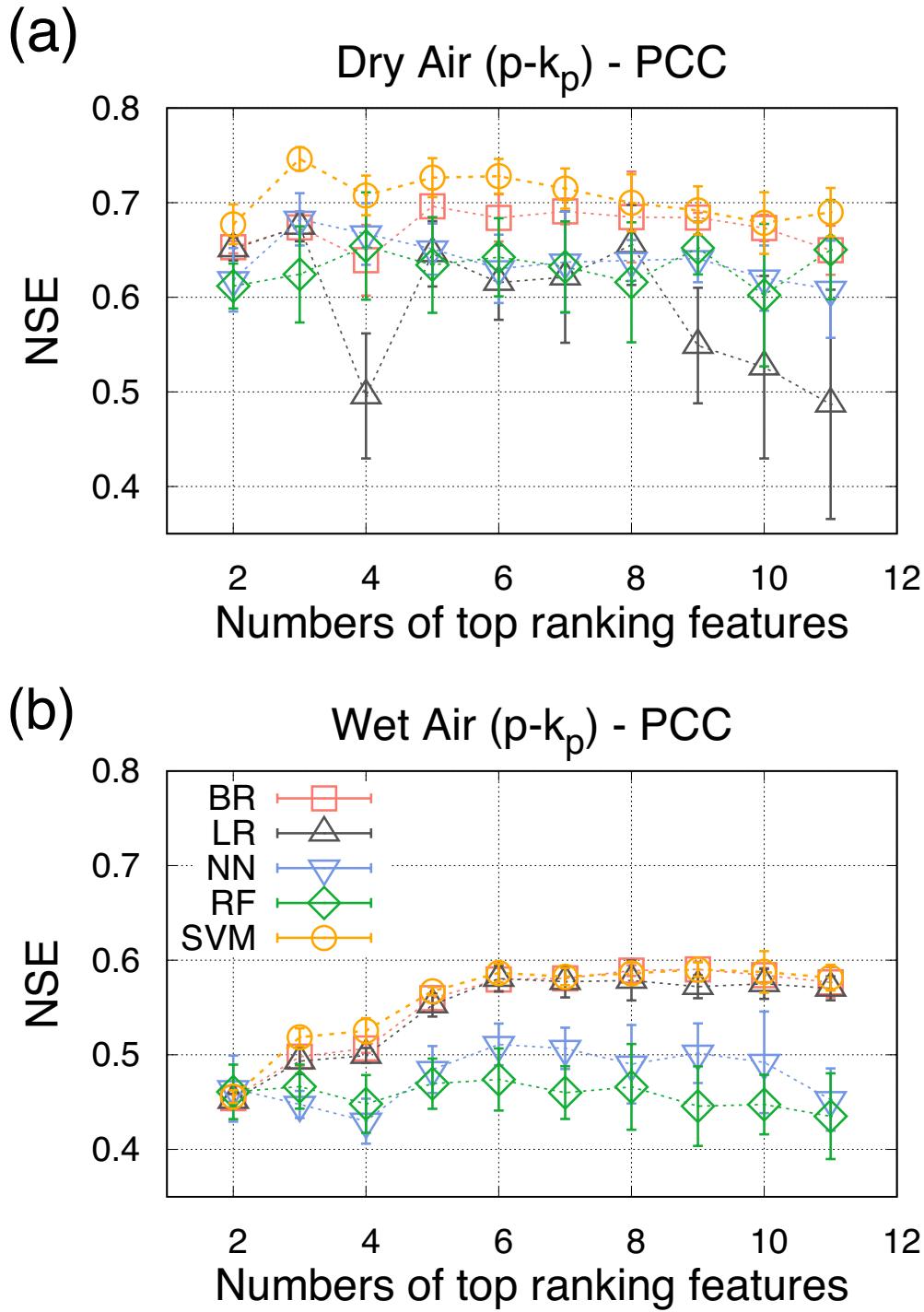

Figure 5 **Performance of trained machine learning models.** NSE of five trained ML models (BR: Bayesian ridge regression, LR: linear regression, NN: nearest neighbor, RF: random forest, and SVM: support vector machines regression) as a function of the number of top-ranking features from the PCC analysis in dry air and wet air.

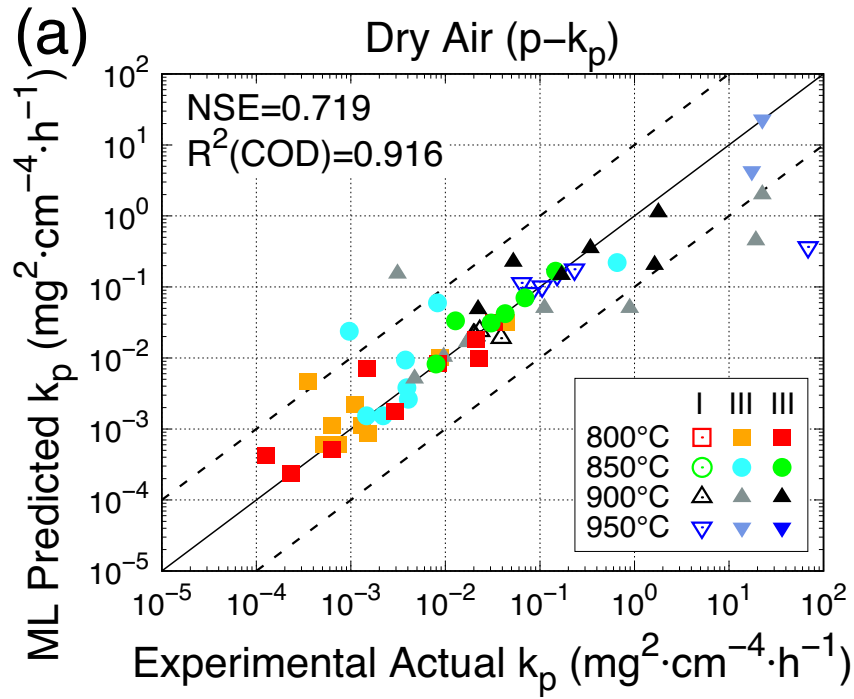

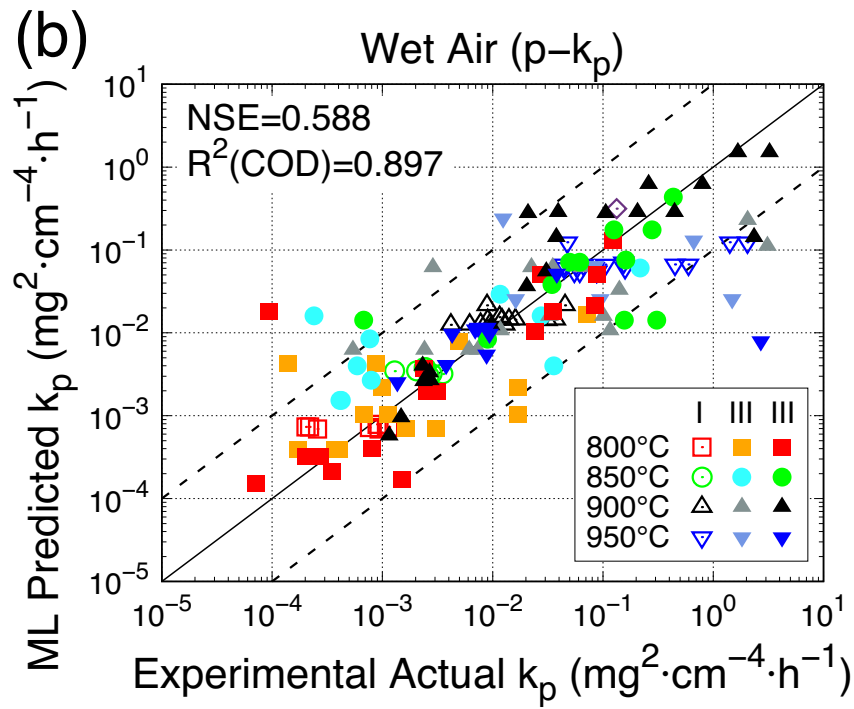

Figure 6 **Machine learning parity plots**. Comparison bewteen experimental and SVM-predicted $k_p$ of three groups of alloys (I: commercial alloys, II: Ni-Cr binary alloys and III: Ni-Cr alloys with additions) at different oxidation temperatures. The region between the dashed lines indicates one order of magnitude fidelity range as an acceptable deviation.

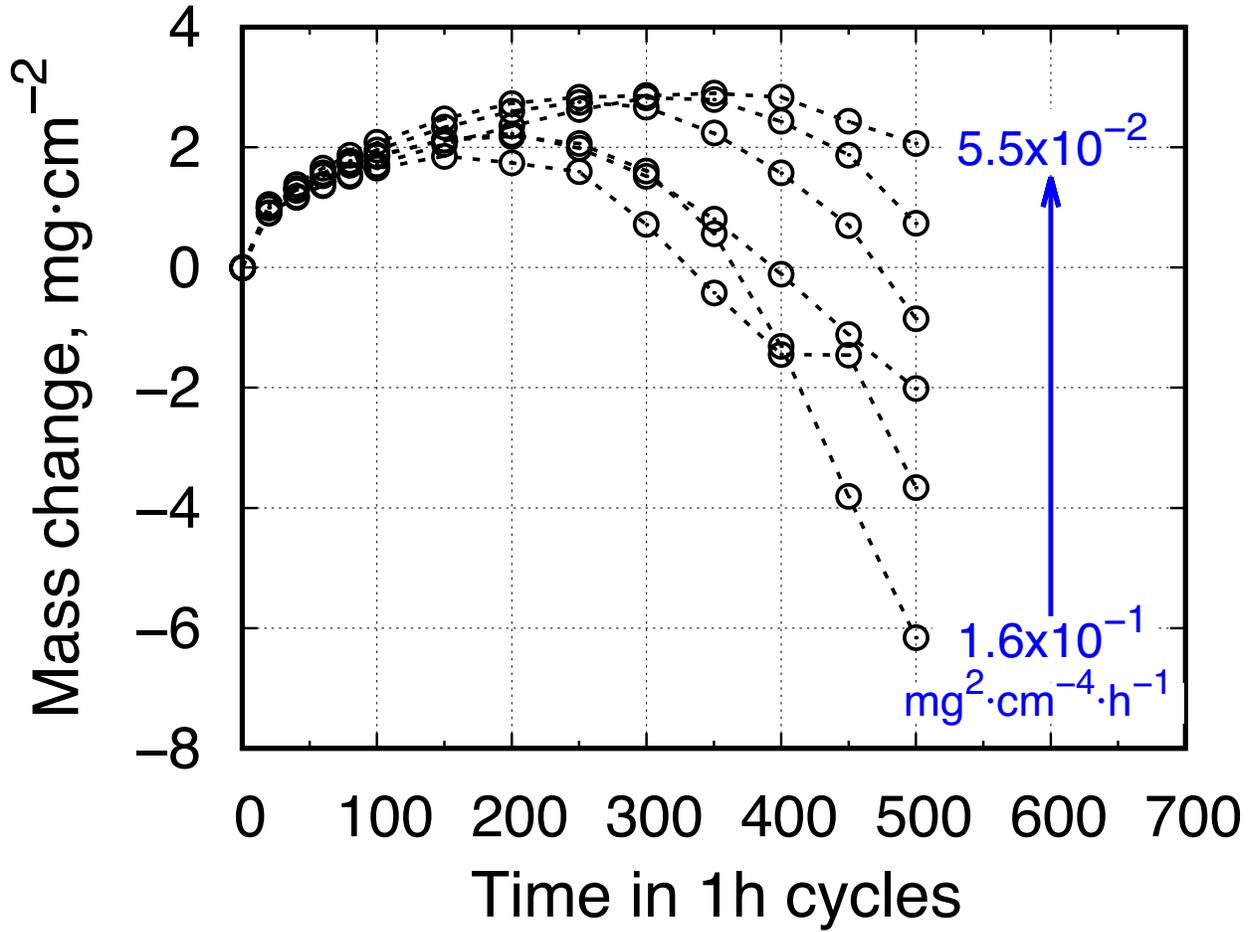

Figure 7 **Oxidation of Nimonic 80A at 950 °C in wet air in 1 hour cycle.** The $k_p$ determined by the $p$-$k_p$ model largely varies from six tests under the same condition, whereas the one from s-$k_p$ does not.

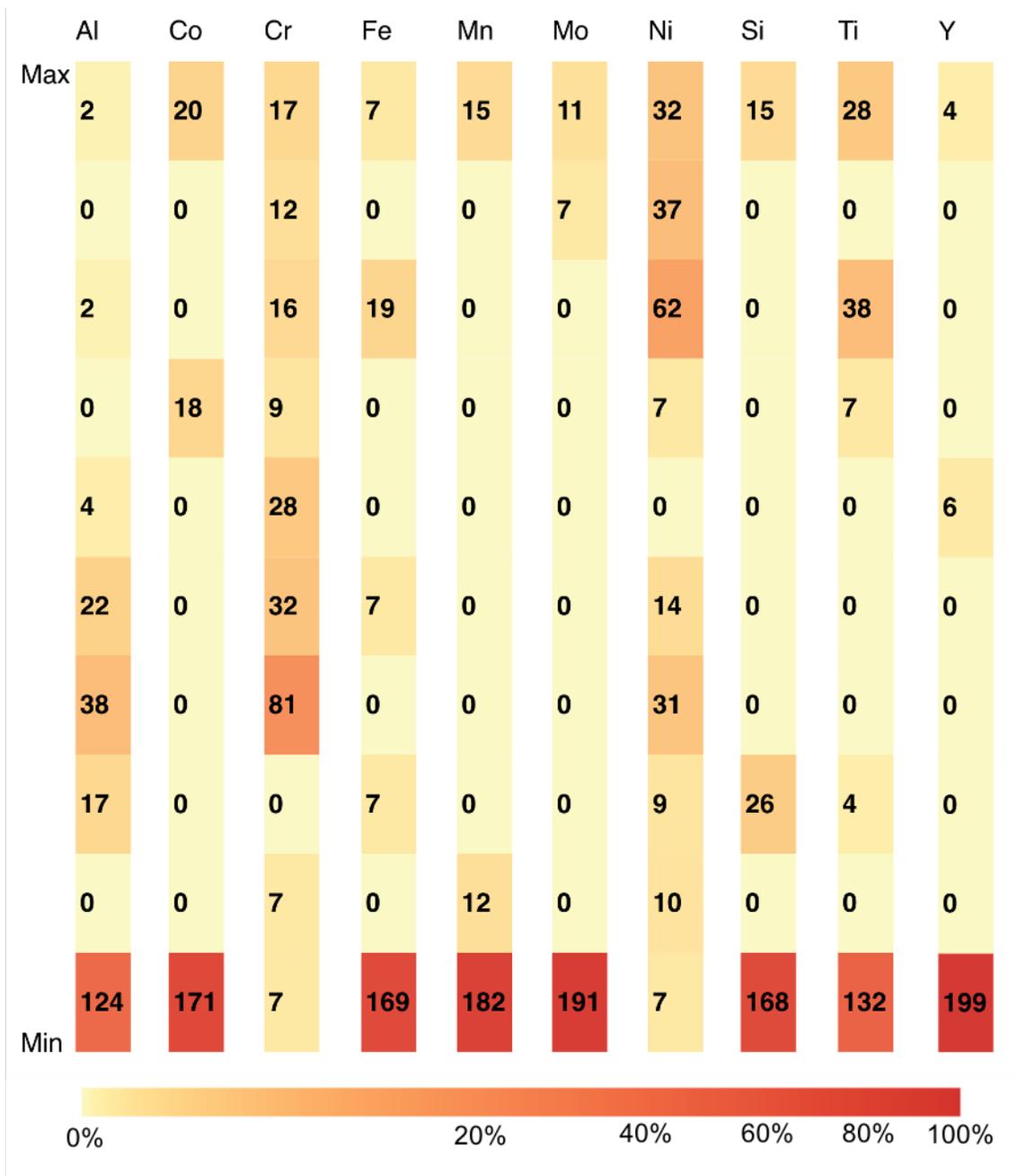

Figure 8 **Distribution of cyclic oxidation data.** The concentration range of each element, i.e., the different between maximum and minimum content, is evenly divided into 10 segments. The number in each segment represents how many data points are located in each cell. For example, a cell with "0" means that there is no data in this range. The degree of data aggregation in each segment is represented by color.

# Supplementary materials for

# Data Analytics Approach to Predict High-Temperature Cyclic Oxidation Kinetics of NiCr-based Alloys

Jian Peng, Rishi Pillai, Marie Romedenne, Bruce A. Pint,

Govindarajan Muralidharan, J. Allen Haynes, Dongwon Shin*

Materials Science and Technology Division, Oak Ridge National Laboratory, Oak Ridge, TN 37831

*Email address: shind@ornl.gov

**Supplementary Note 1**

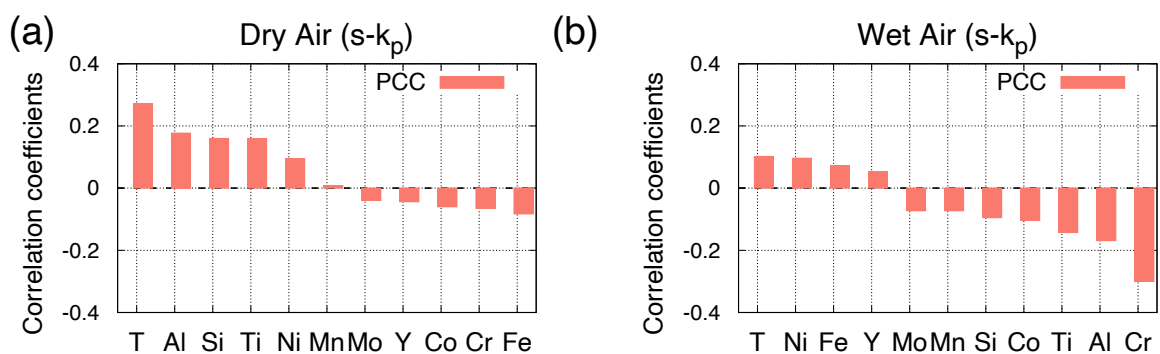

Figure S1 **Correlation analysis.** Quantified correlation scores of input features (elemental compositions, oxidation temperatures) and $k_p$ determined by the s-$k_p$ models, respectively, using Pearson methods at dry air and wet air. Details are reported in the Supplementary Table S1.

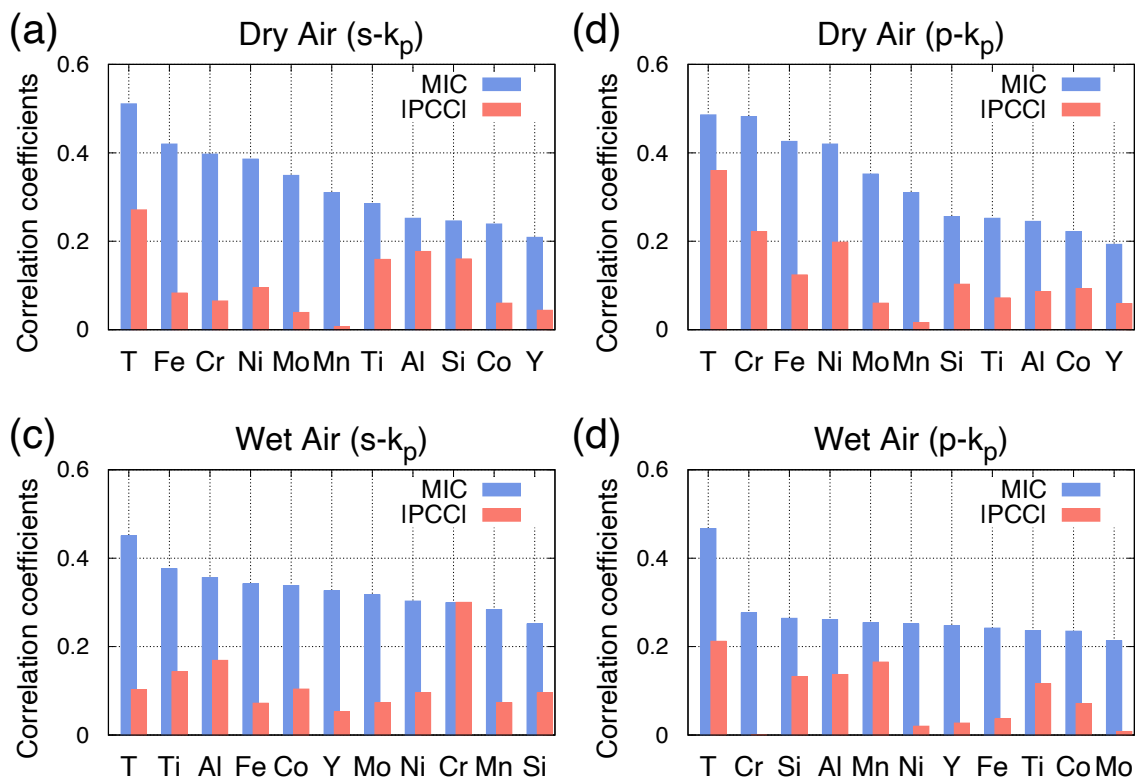

Figure S2 **Correlation analysis.** Quantified correlation scores of input features (elemental compositions and oxidation temperatures) and $k_p$ determined by s-$k_p$ and p-$k_p$ models, respectively, using MIC methods at difference atmpsphere. Details are reported in the Supplementary Table S1. |PCC| denotes the absolute value of PCC coefficients.

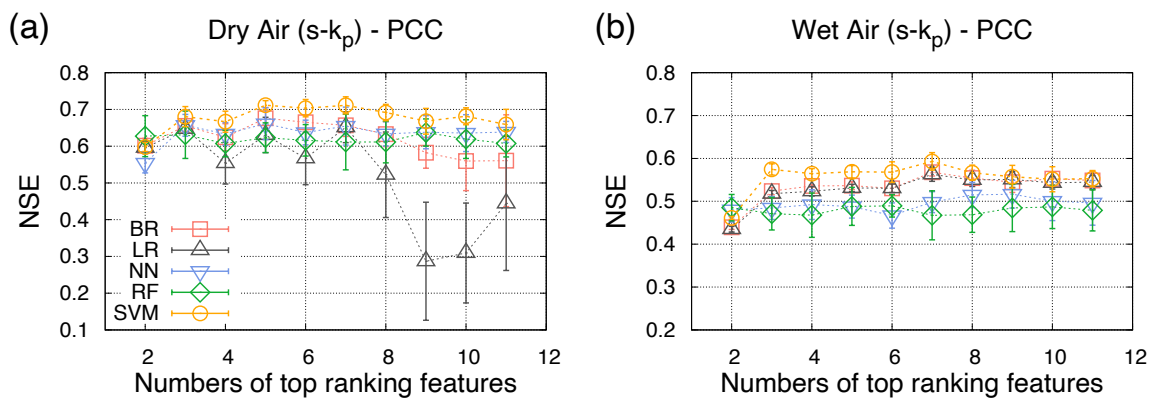

Figure S3 **Performance of trained ML models.** NSE of five trained ML models (BR: Bayesian ridge regression, LR: linear regression, NN: nearest neighbor, RF: random forest, and SVM: support vector machines regression) as a function of the number of top-ranking features from the PCC analysis in dry air and wet air.

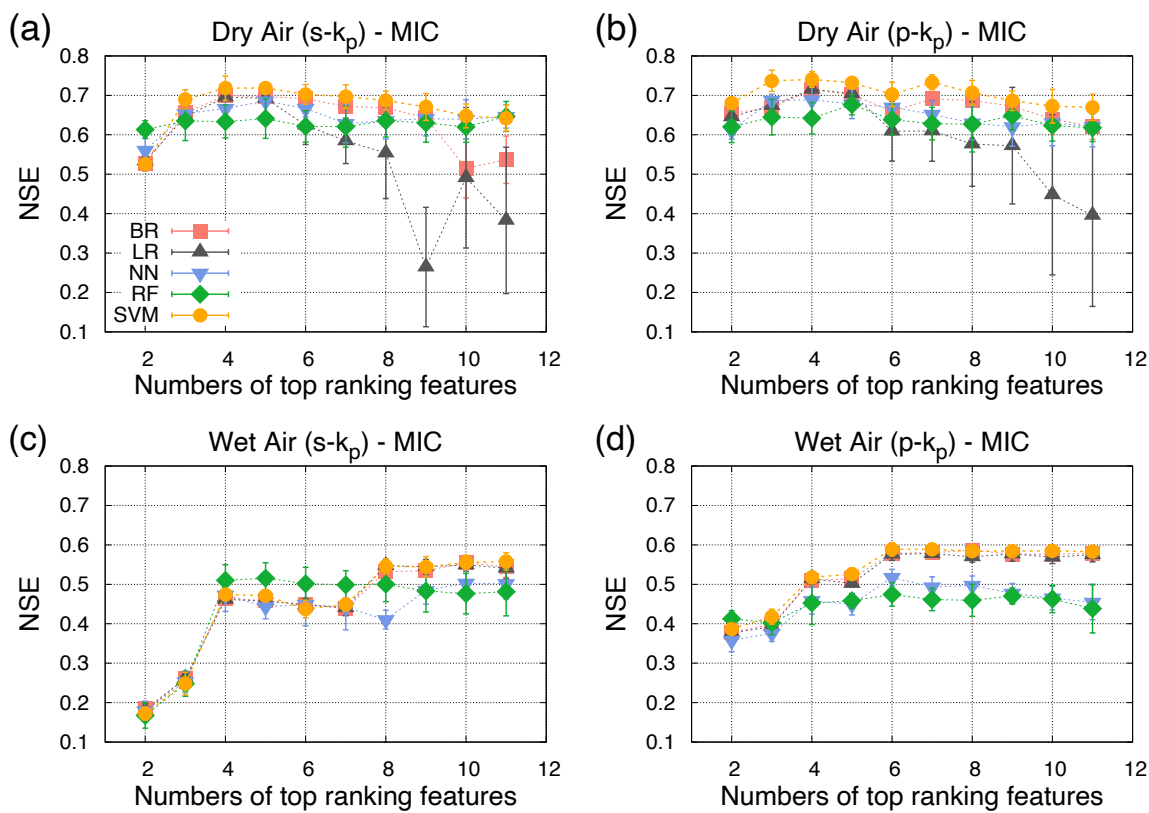

Figure S4 **Performance of trained ML models.** NSE of five trained ML models (BR: Bayesian ridge regression, LR: linear regression, NN: nearest neighbor, RF: random forest, and SVM: support vector machines regression) as a function of the number of top-ranking features from the MIC analysis in dry air and wet air.

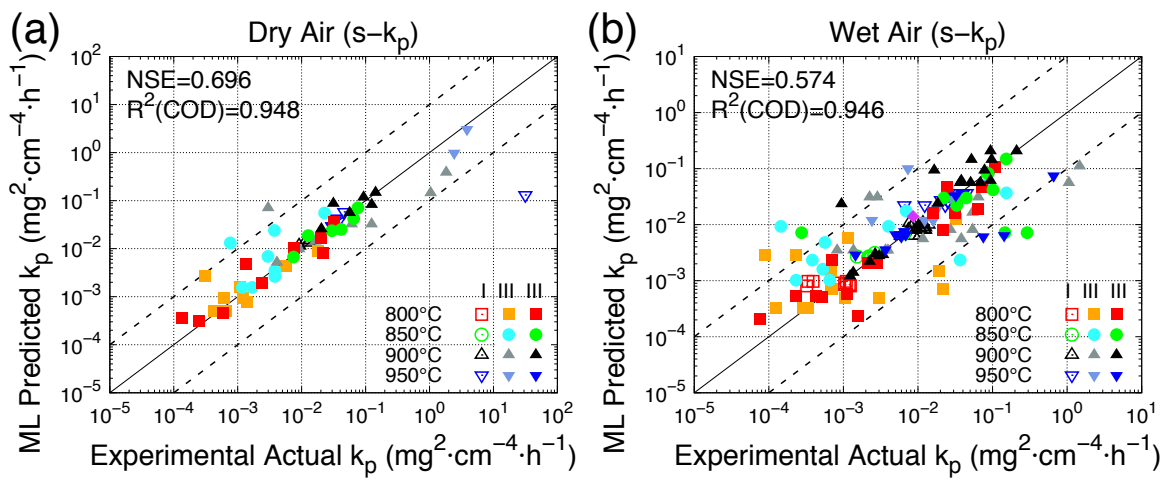

Figure S5 **ML parity plots**. Comparison of fitted and SVM-predicted $k_p$ of three groups of alloys (I: commercial alloys, II: Ni-Cr binary alloys and III: Ni-Cr alloys with additions) at different oxidation temperatures. The region between the dashed lines indicates one order of magnitude fidelity range as the acceptable deviation.

Table S1 Summary of correlation analysis results between input features (elemental compositions and oxidation temperatures) and $k_p$ determined by s-$k_p$ and p-$k_p$ models, respectively, of Ni-Cr-based alloys using MIC and PCC methods. |PCC| denotes that the absolute value of the correlation coefficients of PCC.

|  |  | s_$k_p$ model | | | | | p_$k_p$ model | | | | |
| --- | --- | --- | --- | --- | --- | --- | --- | --- | --- | --- | --- |
|  |  | PCC | |PCC| | |PCC| ranking | MIC | MIC Ranking | | PCC | |PCC| | |PCC| ranking | MIC | MIC Ranking |
| Dry air | T | 0.271 | 0.271 | 1 | 0.511 | 1 | T | 0.36 | 0.36 | 1 | 0.486 | 1 |
|  | Al | 0.177 | 0.177 | 2 | 0.252 | 8 | Ni | 0.198 | 0.198 | 3 | 0.42 | 4 |
|  | Si | 0.16 | 0.16 | 3 | 0.246 | 9 | Si | 0.103 | 0.103 | 5 | 0.256 | 7 |
|  | Ti | 0.159 | 0.159 | 4 | 0.285 | 7 | Al | 0.086 | 0.086 | 7 | 0.245 | 9 |
|  | Ni | 0.095 | 0.095 | 5 | 0.386 | 4 | Ti | 0.072 | 0.072 | 8 | 0.252 | 8 |
|  | Mn | 0.007 | 0.007 | 11 | 0.31 | 6 | Mn | -0.016 | 0.016 | 11 | 0.31 | 6 |
|  | Mo | -0.039 | 0.039 | 10 | 0.349 | 5 | Y | -0.059 | 0.059 | 10 | 0.193 | 11 |
|  | Y | -0.044 | 0.044 | 9 | 0.209 | 11 | Mo | -0.06 | 0.06 | 9 | 0.352 | 5 |
|  | Co | -0.06 | 0.06 | 8 | 0.239 | 10 | Co | -0.093 | 0.093 | 6 | 0.222 | 10 |
|  | Cr | -0.065 | 0.065 | 7 | 0.397 | 3 | Fe | -0.124 | 0.124 | 4 | 0.426 | 3 |
|  | Fe | -0.083 | 0.083 | 6 | 0.42 | 2 | Cr | -0.222 | 0.222 | 2 | 0.482 | 2 |
| Wet air | T | 0.102 | 0.102 | 5 | 0.45 | 1 | T | 0.212 | 0.212 | 1 | 0.467 | 1 |
|  | Ni | 0.096 | 0.096 | 6 | 0.303 | 8 | Mn | 0.165 | 0.165 | 2 | 0.254 | 5 |
|  | Fe | 0.072 | 0.072 | 10 | 0.342 | 4 | Si | 0.132 | 0.132 | 4 | 0.264 | 3 |
|  | Y | 0.052 | 0.052 | 11 | 0.326 | 6 | Fe | 0.036 | 0.036 | 7 | 0.242 | 8 |
|  | Mo | -0.073 | 0.073 | 9 | 0.317 | 7 | Ni | 0.02 | 0.02 | 9 | 0.251 | 6 |
|  | Mn | -0.073 | 0.073 | 8 | 0.283 | 10 | Cr | -0.001 | 0.001 | 11 | 0.276 | 2 |
|  | Si | -0.095 | 0.095 | 7 | 0.251 | 11 | Mo | -0.008 | 0.008 | 10 | 0.213 | 11 |
|  | Co | -0.104 | 0.104 | 4 | 0.338 | 5 | Y | -0.027 | 0.027 | 8 | 0.247 | 7 |
|  | Ti | -0.143 | 0.143 | 3 | 0.376 | 2 | Co | -0.071 | 0.071 | 6 | 0.235 | 10 |
|  | Al | -0.169 | 0.169 | 2 | 0.356 | 3 | Ti | -0.116 | 0.116 | 5 | 0.236 | 9 |
|  | Cr | -0.3 | 0.3 | 1 | 0.299 | 9 | Al | -0.137 | 0.137 | 3 | 0.261 | 4 |